# Retractions in Arts and Humanities: an Analysis of the Retraction Notices

Ivan Heibi [1,2], Silvio Peroni [1,2]

[1] Research Centre for Open Scholarly Metadata, Department of Classical Philology and Italian Studies, University of Bologna, Bologna, Italy

[2] Digital Humanities Advanced Research Centre (/DH.arc), Department of Classical Philology and Italian Studies, University of Bologna, Bologna, Italy

## Abstract

The aim of this work is to understand the retraction phenomenon in the arts and humanities domain through an analysis of the retraction notices – formal documents stating and describing the retraction of a particular publication. The retractions and the corresponding notices are identified using the data provided by Retraction Watch. Our methodology for the analysis combines a metadata analysis and a content analysis (mainly performed using a topic modeling process) of the retraction notices. Considering 343 cases of retraction, we found that many retraction notices are neither identifiable nor findable. In addition, these were not always separated from the original papers, introducing ambiguity in understanding how these notices were perceived by the community (i.e., cited). Also, we noticed that there is no systematic way to write a retraction notice. Indeed, some retraction notices presented a complete discussion of the reasons for retraction, while others tended to be more direct and succinct. We have also reported many notices having similar text while addressing different retractions. We think a further study with a larger collection should be done using the same methodology to confirm and investigate our findings further.

## Introduction

When a scholarly publication is retracted, the organization/entity responsible for publishing it (e.g., a journal) has decided to withdraw it because of errors or other irregularities. Retractions can be either partial or full, with partial retractions involving only small portions of the article being flawed or containing errors. Correcting these flawed portions does not compromise the overall information and conclusions of the article. However, full retractions are necessary when the content or data of an article is so seriously flawed or erroneous that its findings and conclusions cannot be trusted. As defined by Barbour *et al.* (2009), a full retraction is *"a mechanism for correcting the literature and alerting readers to articles that contain such seriously flawed or erroneous content or data that their findings and conclusions cannot be relied upon"*.

According to Vuong *et al.* (2020), most retractions occur in STEM fields, with social sciences and humanities accounting for a relatively small percentage. Generally, the reasons for retraction fall into two main categories: (a) honest mistakes and (b) misconduct. The decision to retract an article rests with the publication's editor(s). It must be accompanied by a formal document called a retraction notice. This notice should contain sufficient details about why the findings are no longer reliable and explicitly state whether the retraction was due to honest error or misconduct. The PDF and online versions of the retracted article should provide free access to the retraction notice, as recommended by Barbour *et al.* (2009).

Several past studies investigated the reasons for retraction. On the one hand, great attention has been given to STEM, such as in health science (Li et al., 2018), anesthesiology (Nair et al., 2020), engineering (Rubbo et al., 2019), computer science (Al-Hidabi and Teh, 2019), economics and management (Karabag and Berggren, 2012). On the other hand, few past studies worked on analysing the arts and humanities domain – an example is the work of Halevi (2020) and our prior work on the topic (Heibi and Peroni, 2022a). Reasons for this disparity in the literature attention in favour of STEM might be related to either their high numbers or the fact that retractions in STEM are more likely to be perceived as potential threats, especially for retracted publications linked to sensitive areas of study, such as medicine (Heibi and Peroni, 2022a).

Considering the less attention that has been given to the study and analysis of the retraction phenomenon in the arts and humanities domain and, in particular, to the reasons for retraction, we present an attempt to address these issues and understand such behaviour through a special focus on the retraction notices of fully retracted articles.

In particular, our work is based on an automatic analysis of the retraction notices considering their related metadata and textual content. The content analysis process is mainly based on the application of a topic modeling analysis on the full text of the retraction notices, performed using MITAO, a visual interface to create a customizable visual workflow for text analysis (Ferri *et al.*, 2020), which was also tested on the analysis of a humanities corpora (Heibi *et al.*, 2020). The designed methodology takes into consideration both quantitative aspects (e.g., number of retractions having a specific retraction notice) and qualitative ones (e.g., through an analysis of the textual structure of the retraction notices, or the most common words used to address a specific reason of retraction). To identify and collect metadata of the retractions, we relied on the dataset provided by Retraction Watch (Oransky and Marcus, 2012), a blog that works on reporting retractions (and related topics) of scientific papers and collect/label their related metadata[1].

The rest of this paper is organized as follows. In the section "Background", we provide a comprehensive review of the literature on how retractions and retraction notices have been analysed in the past and what have been the main results. Section "Methodology" presents our methodology and describes the tools and data sources we used. Section "Analysis and Results" walks through the methodology steps and presents the results of our analysis, while Section "Discussions" discusses the most relevant aspects we have observed when combining the results of our analysis. Finally, in Section "Conclusions", we offer our conclusions and final thoughts.

## Background

Same as for STEM, also arts and humanities domain experiences retractions, although there are notable distinctions. Methodologically, humanities research is heavily dependent on intuition, as opposed to STEM's emphasis on reason and logic. However, certain humanities research adopts scientific methods inspired by STEM disciplines (e.g., digital humanities studies) (Huang and Chang, 2008). Essentially, arts and humanities research uses historical, interpretive, and analytical techniques, in contrast to STEM's emphasis on hard evidence and tabular data to form conclusions. These peculiarities of arts and humanities make them prone to retractions, whereas in STEM retractions are more frequent, as shown by Vuong *et al.* (2020).

---

[1] Data available from The Center For Scientific Integrity, the parent non-profit organization of Retraction Watch, subject to a standard data use agreement.

The study of Halevi (2020) is a rare work on a retraction analysis of the arts and humanities domain, and was based on the annotations/data provided by Retraction Watch. The outcomes of this work revealed that the most recurring reason for retraction in arts and humanities is "significant overlap with previously published research" and "plagiarism" (representing 77% of the total). The detection and the forms of plagiarism are well-defined and less prone to interpretation (Dhammi and Ul Haq, 2016), which might explain these results. These results have been confirmed by our work (Heibi and Peroni, 2022a). As for Halevi (2020), our analysis considered the reasons for retraction annotated by Retraction Watch: almost 23% of the retractions in arts and humanities have been classified with the reason "plagiarism of article". In addition, the reported numbers on retraction in arts and humanities indicate an increasing trend throughout the years, which increases the importance of studying this phenomenon.

One of the main aims of this paper is to deepen the analysis of the retraction reasons in the arts and humanities. One way to do that is to consider and automatically analyse the text of the retraction notices to get more details and information regarding the retracted publication addressed by the retraction notice.

Other works, although different domains (or not explicitly domain-based), have considered the retraction notices in their analysis. The work done by Moylan and Kowalczuk (2016) presented an investigation about retractions from BioMed Central journals through an analysis of the retraction notices content. Vuong (2020) presented the limitations of retraction notices (based on a random sample of 2,046 retracted papers) and discovered that nearly 10% either omit or do not contain information related to reasons for retractions. The lack of information given in the retraction notices was also observed by Chambers *et al.* (2019) in their analysis of the reasons for retractions in obstetrics and gynaecology articles: in many cases, no specific reason has been even provided, and others lacked pertinent details. Others considered the retraction notices of specific individual retraction cases they analysed in their papers (Schneider et al., 2020; Luwel et al., 2019; Heibi and Peroni, 2021). To the best of our knowledge, approach we followed in the present work represents a first attempt to deeper understand retraction in humanities through a special focus on retraction notices.

## Methodology

As summarized in Fig. 1, our methodology is articulated in four main steps. The design of this methodology takes inspiration from our experience in the analysis of retractions (Heibi and Peroni, 2021; 2022a), and it is based (in the parts related to the use case of this paper) on the methodological protocol we developed for gathering, characterizing, and analysing incoming citations of retracted articles (Heibi and Peroni, 2022b).

First, we need to identify and collect all the retraction notices of all the retracted publications in arts and humanities. We do not establish specific criteria for identifying items in the arts and humanities; instead, we rely on the classification provided by Retraction Watch. The dataset provided by Retraction Watch Database (http://retractiondatabase.org/), that stores retractions with their corresponding metadata following the reports on retractions made by Retraction Watch (Oransky and Marcus, 2012).

Then the data analysis is split into two parts: (A1) content analysis and (A2) metadata analysis. On the one hand, the content analysis is based on two sub-steps: (A1.1) collecting the full text of the retraction notices, and (A1.2) building and running a topic modelling (TM) analysis using the Latent Dirichlet Allocation (LDA) model (Jelodar et al., 2019). On the other hand, the metadata analysis is based on: (A2.1) annotating relevant metadata using the dataset of Retraction Watch, and (A2.2) annotating additional metadata using external sources. Finally, we combine the content analysis results and the metadata values to infer quantitative and qualitative findings to help us understand the retraction nature/behaviour in the arts and humanities domain.

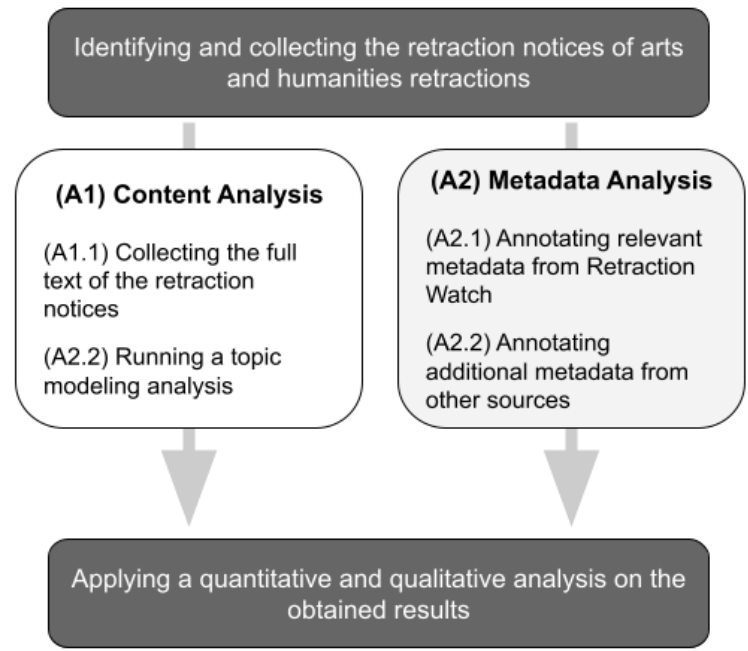

**Figure 1**. A schema representing the adopted methodology. First, we identify/collect the retraction notices in the arts and humanities domain using the Retraction Watch dataset, and then the analysis is made in two parts. On the one side, we apply a content analysis (A1), while, on the other side, we perform a metadata analysis (A2). Finally, the outcomes of (A1) and (A2) are combined and discussed.

## Analysis and Results

In this section, we discuss, analyse, process, and describe the results obtained by each step of our methodology. Each step is treated and presented in a separate section.

### *Identifying the retraction notices*

First, to gather the retraction notices, the actual retractions themselves need to be identified. One possible approach would have been querying large bibliographic databases, such as ScienceDirect by Elsevier, searching for terms such as "RETRACTED". Yet, in this work, we decided to use the Retraction Watch Database to identify all the retracted publications that fall under the arts and humanities domain. The database reports on retractions of scientific papers and curates the related metadata of reported retractions. One main reason which guided us through this decision was that the metadata of each retracted paper included its corresponding field of study; therefore, this helped us to filter the dataset and gather only retractions related to the arts and humanities domain.

Retraction Watch provided the dataset used in this work after signing a data use agreement. The analysed dataset was based on the data stored by Retraction Watch up to May 2023.

The dataset provided by Retraction Watch specifies a column "Subject", defining the corresponding field(s) of study of a given retracted publication (row). Retraction Watch assigns to each retracted publication in the dataset a macro area (e.g., "HUM" for humanities) and a category of study (e.g., "Arts – Literature/Poetry"). One retracted publication can have multiple areas/categories of study. For instance, a retracted article $R$ in the dataset may be labelled with the areas of study "*(HUM) Arts - Literature/Poetry*" and "*(SOC) Philosophy*". In this case, $R$ is assigned to the two macro areas HUM (i.e., arts and humanities) and SOC (i.e., social sciences) and respectively to the categories "*Arts - Literature/Poetry*" and "*Philosophy*". Retraction Watch encompasses 18 subjects in the arts and humanities macro area, including history, philosophy, religion, and more.

The total number of retracted publications having at least one of their subjects in the arts and humanities domain is 1,167. To reduce the bias in the final results, we considered only the retracted publications with all their subjects under the HUM macro area. This collection is composed of 343 publications.

## *Content Analysis*

The content analysis of the retraction notices (i.e. A1) is based on two steps: (A1.1) collecting the full text of the retraction notices and (A1.2) running a topic modeling analysis.

First, we gathered the retraction notices of the retracted publications identified in the previous phase (filtered to contain only retracted publications in arts and humanities). Retraction Watch data included the DOI (if any) for each retraction notice in the dataset. We considered only the retraction notices that were identifiable and findable. The total number of retracted publications with a corresponding identifiable retraction notice was 218 (out of 343).

Of these 218 retraction notices, we did not find 10 of them, although they had a DOI specified. In almost all these cases, the linked resource was the original retracted paper which did not report any retraction notice. In addition, we found 3 additional inaccessible retraction notices due to paywall restrictions. All these cases (13) have been excluded from the collection, which finally contained 205 retraction notices.

The relation between the retracted publications and their corresponding retraction notices was not one-to-one (i.e., each retracted publication has a different and exclusive retraction notice). Indeed, some retracted publications may have been linked to the same retraction notice, which might address separately each retracted publication in its content. We found 7 retraction notices linked to multiple publications. In particular, we reported retraction notices assigned to 13, 4, and 3 retracted publications, respectively, and 2 retraction notices each assigned to 7 and to 2 different retracted papers, respectively. Following these considerations, we had 38 duplicate retraction (13 + 4 + 3 + (7 * 2) + (2 * 2)). Therefore, the final number of retraction notices we considered was 174.

We gathered the full text of all 174 retraction notices. This collection represents the corpus we used for the content analysis step. Such a step was mainly based on the application of a topic modeling analysis.

A topic modeling analysis is a statistical approach for automatically discovering the topics (represented as a set of words) that occur in a collection of documents. In other words, the process analyse texts and creates "topics", which are bags of words that often co-occur together in the original texts (Mohr and Bogdanov, 2013). Topic modeling has been used in several domains. In the arts and humanities domain, for instance, to study the different uses of language in academia (McFarland et al., 2013), or for analysing the themes introduced in about 3,200 19th-century novels (Jockers and Mimno, 2013). The topic modeling analysis was crucial in many of our previous studies regarding the analysis of retractions (Heibi and Peroni, 2021, 2022a), and it represents one of the main phases of the protocol we have designed to deal with these studies (Heibi and Peroni, 2022b).

To perform the topic modeling, we used the tool suggested in (Heibi and Peroni, 2022b), i.e., MITAO – Mashup Interface for Text Analysis Operations (Ferri et al., 2020). MITAO makes topic modelling techniques usable by scholars with no or limited coding skills via a visual interface which enables them to easily create a visual workflow (which hides technical specifications and coding) for processing textual content. One of the main features of MITAO, which led us to this choice, is that it provides a mechanism for sharing the adopted workflow with the research community to foster the reproducibility of the process, which is a fundamental aspect we wanted to guarantee for our research. In addition, the workflow (once shared) can be modifiable by others that lack programming skills or the ability to understand programming languages to a certain extent. Finally, MITAO was also successfully adopted for an analysis of an arts and humanities use case (Heibi *et al.*, 2021).

Using MITAO, we built the topic model using the Latent Dirichlet Allocation (LDA) model (Jelodar et al., 2019). A standard topic modeling workflow is usually based on three main steps: tokenization, vectorization, and topic model creation. In our work, one of the input of this process was a corpus with the full texts of the retraction notices we gathered. Another crucial input, though, was the number of topics to retrieve, which needed to be determined in advance. The approach we decided to adopt in this work was based on the value of the topic coherence score, as suggested in the work of Schmiedel *et al.* (2019). The coherence score is used to measure the degree of the semantic similarity between high-scoring words in the topic. It helps compare topics that are semantically interpretable from the topics that are artefacts of a mere statistical inference. Using this approach, we calculated the average coherence score for a range of topic models trained with a different number of topics and following the obtained results, we decided to train our topic model with 3 topics and to use the LDAvis (Sievert and Shirley, 2014) for showing them, which integrated into MITAO.

LDAvis plots the topics as circles in a two-dimensional plane whose centers are determined by computing the distance between the topics and uses multidimensional scaling to project the inter-topic distances onto two dimensions. The dimension of the area of each circle represents the topic prevalence. A list of 30 terms ranked using the term saliency measure is displayed. This saliency measure combines the overall probability of a term with its distinctiveness: how informative is a specific term for determining the generation of a topic versus any other randomly selected term. By selecting a particular topic, LDAvis shows a list of 30 terms ranked using the relevancy measure, which ranks terms within each topic to aid users in topic interpretation activities. This measure is controlled by a weight parameter $\lambda$, which allows one to rank either the terms in decreasing order of their topic-specific probability if close to 1 or terms solely by their lift if close to 0.

We tuned LDAvis by adjusting the relevance metric as recommended by Sievert and Shirley (2014) to produce the graphs in Fig. 2, that shows the 3 topics identified.

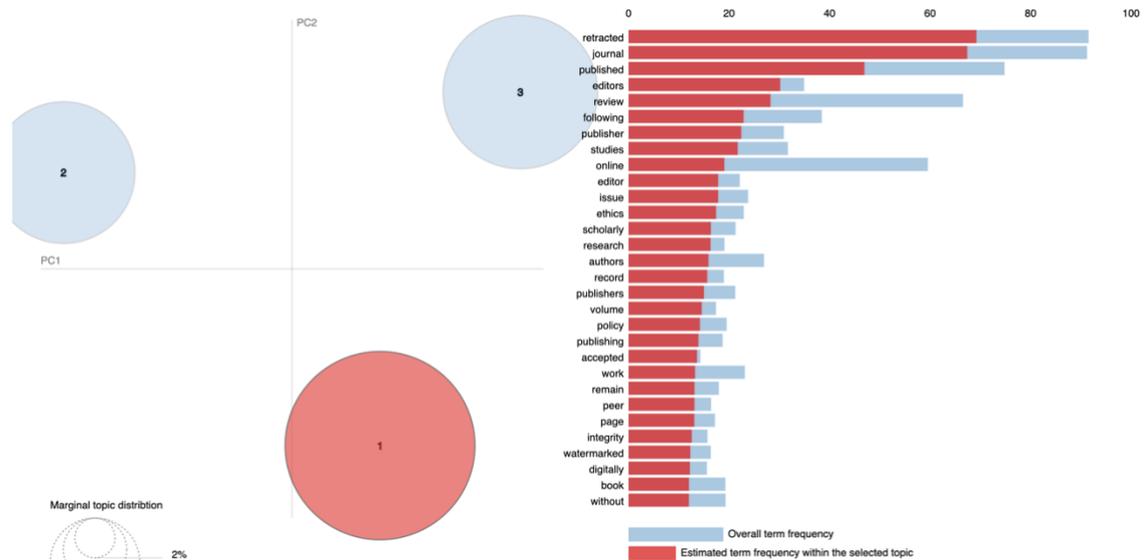

**Figure 2.** The three topics visualized using LDAvis, where Topic 1 is selected. This visualization is available at https://ivanhb.it/hum-ret-notices-results/data/ldavis.

The five most common terms are: "retracted", "journal", "published", "editors", and "content". The largest topic of the three was Topic 1 and included all these terms (it contained 44.8% tokens overall). In addition, Topic 1 contained other terms explicitly related to the retraction phenomenon, such as "plagiarism", "integrity", "ethics", and "cope". Topics 2 and 3 included common terms used in the actual content of the notices, such as "violation", "found", "remove", "problem", and "apologize". Topic 2 included additional terms such as "ieee" and other names referring to organizations and people, that are often included in the content of the retraction notices.

The outcomes produced are available at https://ivanhb.it/hum-ret-notices-results/ for a dynamic consultation. Raw data are published on Zenodo (Heibi, 2023).

## Metadata Analysis

To characterize the retraction notices gathered with some main metadata, we relied on and queried only open repositories storing scholarly bibliographic data compliant with the Open Science principles. In particular, we decided to rely on the bibliographic and citation data freely provided (under CC0 licence) by OpenCitations (http://opencitations.net/) (Peroni and Shotton, 2020).

Starting from the data provided by Retraction Watch, we retrieved and annotated some basic metadata for each retraction notice and its corresponding retracted publication (343). In this case we considered all the collection including those that are not identifiable (which was essential in the content analysis). More precisely, the metadata we considered were the subject (e.g. "History"), journal, publisher (e.g., "Taylor and Francis"), type of the corresponding retracted publication (e.g., "Journal Article"), date of retraction, date of publication of the corresponding retracted publication, and the reason of retraction.

In addition, we annotated the total number of citations to each retraction notice using COCI, the OpenCitations Index of Crossref open DOI-to-DOI citations (https://opencitations.net/index/coci) (Heibi et al., 2019). We gathered this information usingthe REST API service of OpenCitations (https://opencitations.net/index/api/v1) (Daquino et al., 2022). The service is queryable by giving as input the DOI of the retraction notice of interest – of course, we could not retrieve the citation count of unavailable retraction notices.

The APIs were queried on the 5$^{th}$ of May 2023. In this case, this annotation was possible only to the identifiable notices.

While we successfully collected the metadata of all the retraction notices in the collection, we found only 24 retraction notices with one or more citations. The highest number of citations we have reported to a specific retraction notice was 42.

## Discussions

The total number of retracted publications we gathered was 343. Yet, as discussed in Section "Methodology", only 218 have an identifiable retraction notice. To deepen more on this aspect, we checked the relation between the retractions and their corresponding publishers (shown in Fig. 3). We reported the percentage of retractions which do not have an identifiable retraction notice for each different publisher (the bar beside each publisher in Fig. 3). The results seem to suggest that, even for large important publishers such as Taylor and Francis and Wiley, the retraction notices have not been findable in some cases, which contrasts with the recommendations of the Committee on Publication Ethics (Barbour et al., 2009).

It is important to notice that, while retractions may sometimes be issued either jointly or on behalf of the publisher, the primary responsibility for the journal's content lies with the editor. The editor should always retain the ultimate authority in deciding whether to retract a publication (Barbour et al., 2009). In essence, these findings should consider this factor and provide a balanced perspective on the responsibilities of publishers.

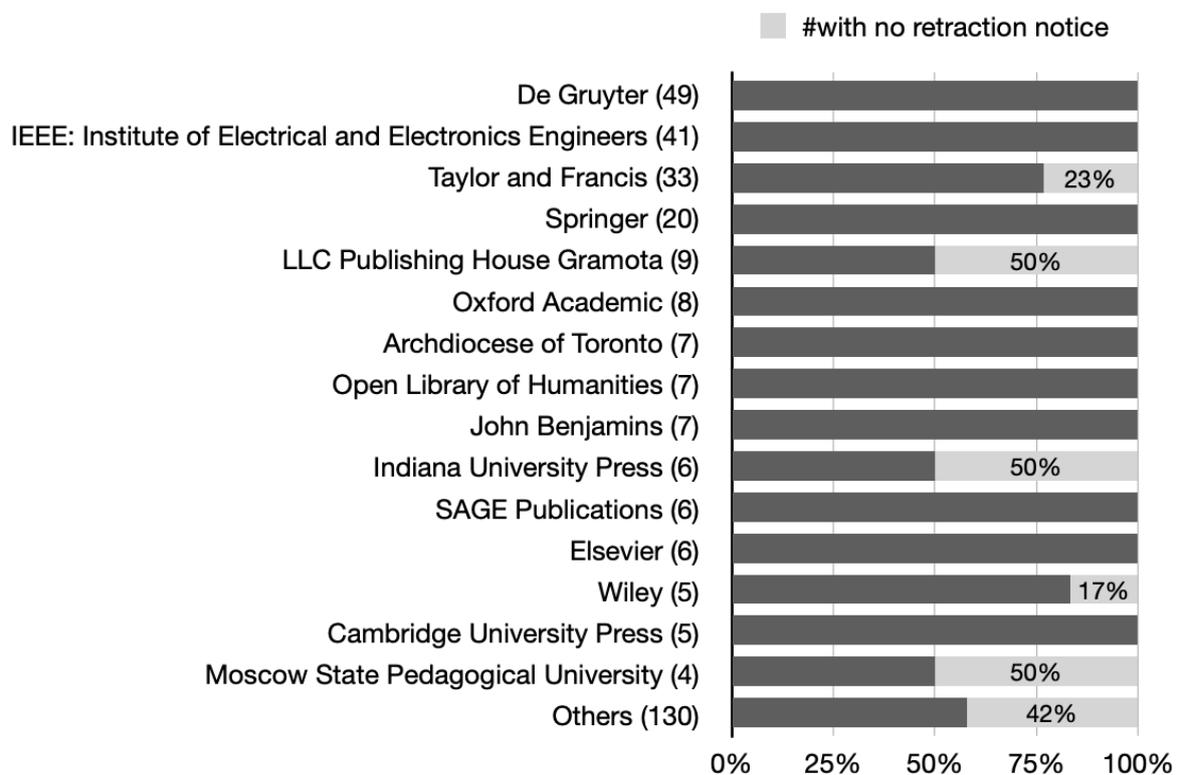

**Figure 3.** The number of retractions in the arts and humanities domain for each different publisher. The bar beside each publisher shows the percentage of retractions without an identifiable retraction notice.

Another aspect we have observed is the fact that 15 retracted publications have an identifiable retraction notice, yet they are not identifiable, i.e., do not have a corresponding DOI. For 122 cases, the retraction notice and the original paper have the same DOI (i.e. the retraction notice has not been published separately). Out of the 24 retraction notices which have received one or more citations, only 6 retraction notices have their own DOI that is different from that of the papers the notices refer to. These retraction notices have received very few citations overall, averaging 1.4 citations per notice. It is important to highlight the fact that, while in this last case, we are sure that the citations are explicitly referring to the retraction notices of corresponding publications, when the original publications and their retractions have the same DOI it is hard to tell whether these citations have been made to the article itself or to refer to the related retraction notice. A further analysis of the context of the in-text reference pointers of such citations is needed to have more insights regarding this aspect.

When analysing the topic modeling results and the most relevant topic of each publisher, we discovered that Topic 2 was highly related to the publisher IEEE. The content of its corresponding retraction notices and the explanation defining the retraction was limited and very much similar (from a textual point of view) in almost all the retraction notices analysed. Another publisher to which we have observed a similar behaviour was De Gruyter. In both cases, the retraction notices discussed the reasons/nature of the retraction with general claims such as *"...this paper has been found to be in violation of IEEE's Publication Principles ..."* or *"... Due to an unforeseeable rights problem, we ..."*.

Generally, from the content analysis of the retraction notices, we did not find a precise, systematic template adopted to describe the retraction. Some retraction notices address the reasons for retraction directly and briefly (even with one sentence, e.g., *"this article was retracted due to translation plagiarism"*). In contrast, others present a larger discussion around the publication to retract. Further analysis should be made to affirm whether this phenomenon is common in other fields of study and what factors may determine this behaviour.

The topics gathered via the topic modeling process reflect the reasons for retraction annotated by Retraction Watch. Indeed, the most common reasons for retraction reported by Retraction Watch to our collection, such as "Copyright Claims", "Plagiarism of Article", and "Duplication of Article", have all been reflected in the terms contained in the generated topics, e.g., "plagiarism", "integrity", and "rights". We believe that a larger corpus should be considered to deepen into a more detailed comparison between our results and the annotations of Retraction Watch,. In addition, other the text of retraction notices should be filtered before applying a topic modeling analysis, to reduce the noise in the obtained results – e.g., removing the name of the publishers such as "ieee", being it very frequent in our topic model.

## Conclusions

In this paper, we investigated the retraction phenomenon in the arts and humanities domain through a special focus on the retraction notices – a formal document which states the retraction and its reasons for a specific publication. We applied a content and metadata analysis over the retraction notices identified and collected using the database provided by Retraction Watch. The total number of retracted publications we have considered throughout our analysis is 343. This collection includes all the retractions marked by Retraction Watch only with subjects related to the arts and humanities domain. The results we obtained are published on Zenodo (Heibi, 2023) and available to be dynamically queried at https://ivanhb.github.io/hum-ret-notices-results.

By studying the results, we investigated and discussed some important aspects. We worked on understanding the relation between having a corresponding identifiable retraction notice and other contextual information, such as the publisher, and we discovered that only 218 retracted publications had an identifiable retraction notice. The citation counts we reported helped us infer other interesting aspects, such as the fact that retraction notices generally receive very few citations when such notices are separated (i.e., have a different DOI) from the original retracted publication – a behaviour that we expected. Finally, the content analysis of the retraction notices helped us infer other interesting qualitative aspects. Indeed, there is no systematic way to write a retraction notice: some present a complete discussion on the reasons for retraction, while others tend to be brief, direct and with less detail. In addition, we reported a relatively large number of retraction notices having very similar text that addressed different publications. Usually, these notices discussed the retraction of the publications from a very abstract and generic point of view, without precise details related to the specific publication.

Since the following work has been done on a few hundred of cases of retraction (i.e., 343), we consider it a pilot and an important first approach toward understanding retractions and the nature of the retraction notices in the arts and humanities domain. Yet, a larger study should be conducted to confirm and investigate our findings deeper. Indeed, extending the collection of retracted publications to all the other retractions assigned in Retraction Watch to subjects in the arts and humanities domain alongside other non-humanities subjects (i.e., a total of 1,167) could be beneficial and will be addressed in future investigations. Indeed, this aspect is crucial, especially when performing a topic modeling analysis, which will be conducted on a larger corpus to help us to tune and limit possible bias.